\documentclass[useAMS,usenatbib]{mn2e}
\usepackage{float,graphics,epsfig,array,amsmath,amssymb}
\usepackage{times}

\title
[The absorption-dominated model for AGN X-ray spectra]
{The absorption-dominated model for the X-ray spectra of 
type\,I active galaxies: MCG--6-30-15}

\author
[L.~Miller, T.~J.~Turner, J.~N.~Reeves]
{L.~Miller$^{1}$,
T.~J.~Turner$^{2,3}$
and J.~N.~Reeves$^{4}$.\\
$^{1}$Dept. of Physics, Oxford University, 
Denys Wilkinson Building, Keble Road, Oxford OX1 3RH, U.K.\\
$^{2}$Dept. of Physics, University of Maryland Baltimore County, Baltimore, MD 21250, U.S.A.\\
$^{3}$Astrophysics Science Division, NASA/GSFC, Greenbelt, MD 20771, U.S.A.\\
$^{4}$Astrophysics Group, School of Physical and Geographical Sciences, 
Keele University, Keele, Staffordshire ST5 8EH, U.K.
}

\begin{document}

\pagerange{\pageref{firstpage}--\pageref{lastpage}} \pubyear{2009}

\maketitle

\label{firstpage}

\begin{abstract}
MCG--6-30-15 is the archetypal example of a type\,I active galaxy
showing broad ``red-wing'' emission in its X-ray spectrum at
energies below the 6.4\,keV\,Fe\,K$\alpha$ emission line and a continuum
excess above 20\,keV.  \citet{miller08a} showed that
these spectral features could be caused by clumpy absorbing material,
but \citet{reynolds09a} have argued that the observed Fe\,K$\alpha$
line luminosity is inconsistent with this explanation unless the
global covering factor of the absorber(s) is very low.  However, the
\citeauthor{reynolds09a} calculation effectively considers the only
source of opacity to be the Fe\,K bound-free transition and neglects
the opacity at the line energy:
correction to realistic opacity decreases the predicted line flux
by a large factor.
We also discuss the interpretation of the
covering factor and the possible effect of occultation by the
accretion disk.  Finally, we consider a model for MCG--6-30-15
dominated by
clumpy absorption, which is consistent with global covering factor 0.45,
although models that include the effects of Compton scattering are
required to reach a full understanding.
Variations in covering fraction may dominate the
observed X-ray spectral variability.
\end{abstract}


\section{Introduction}
MCG--6-30-15 is a $z=0.00775$ type\,I active galaxy that was one of the
first discovered to have in its X-ray spectrum a
broad ``red wing'' of emission below the 
6.4\,keV\,Fe\,K$\alpha$ line \citep{tanaka95a}. This has been interpreted
as reflected emission from an accretion
disk, occurring so close to the black hole that relativistic
effects redshift the Fe\,K$\alpha$ line over a wide observed energy range
\citep[e.g.][]{iwasawa96a,wilms01a,fabian03a}, an effect often known as ``blurring''.
Its significance is that it may allow us to observe 
accretion close to the black hole and even infer the black hole
angular momentum, if highly-redshifted emission is measured 
\citep[e.g.][]{brenneman06a}.

However, the X-ray spectra may also be interpreted as being dominated
by the effects of absorption by material further away from the black hole,
albeit still at interesting radii \citep[e.g.][]{inoue03a}. Recently, 
\citet*[][hereafter MTR]{miller08a} have shown that the substantial X-ray dataset
for MCG--6-30-15, comprising {\em Chandra} HETG, {\em XMM-Newton}
EPIC-pn and RGS and {\em Suzaku} XIS and HXD-PIN observations,
between them covering the range 0.5--40\,keV, could be fitted by an 
absorption-dominated model with no relativistically-blurred emission.  
MTR's analysis remains the only one to date to systematically analyse this 
multi-observatory dataset across the whole available spectral range and modelling the 
observed range of spectral variability.
The model described well
that spectral variability, from a low, more absorbed-looking state to
a higher, less absorbed state, as well as the absorption
lines observed in the grating data.  Such states are common in type\,I AGN
\citep[e.g.][]{miller07a}.
A key ingredient was that at least
one of the absorbing zones should be ``partial-covering'', 
implying the absorber is clumpy.
Some of those zones are 
outflowing \citep{lee01a, young05a} and MTR suggested the absorption is part
of an accretion disk wind, enabling us to
study the complex outflowing winds that are expected from AGN with 
high Eddington ratios \citep{king03a}.  The most absorbing zone in the MTR
model is responsible for creating the excess seen in the {\em Suzaku} data above 20\,keV.

As the absorbing gas is expected to be photoionised, we should see recombination
and fluorescent emission, and in principle we can use its
luminosity to constrain the global covering factor
(i.e. the fraction of the sky covered by the absorber as seen from the ionising
source) - see \citet{turner09a} for a review.
A low line luminosity
might imply that the covering factor is low and that our view 
through an absorber is an unusual sightline.
\citet[][hereafter R09]{reynolds09a} have tried to estimate the expected Fe\,K$\alpha$ fluorescent
line luminosity expected from an absorption model for MCG--6-30-15
and have argued that the implied global covering factor is in the range 0.0175--0.035.
They infer that we are unlikely to be on
a sightline through the absorber, and argue that this model is disfavoured
compared with the relativistically-blurred interpretation of the X-ray spectra.

In this letter we show that the calculation of R09 makes poor
assumptions that invalidate their conclusion and we discuss additional
effects that may change the interpretation of such analysis.
Also, R09 chose not to test the actual model described by MTR, so here we discuss
constraints on the global covering factor in that model. 
MTR supposed the hard excess to arise from a combination of absorption
and reflection, but the composition of this component is
not well constrained by the data.
A number of AGN have recently been discovered to have
hard X-ray excesses that appear best explained by clumpy absorption alone
\citep[e.g.][]{turner09b}, 
so here we also test a model, suggested by
MTR but not tested at that time, in which the spectrum and its variability
are shaped by such high-opacity, clumpy absorption.

\section{Fluorescent line emission from absorbing zones}\label{sec:emission}
\begin{figure}
\begin{center}
\resizebox{60mm}{!}{
\rotatebox{270}{
\includegraphics{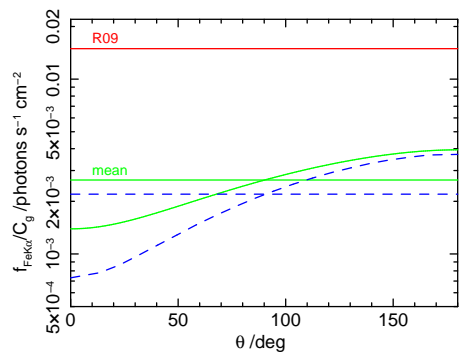}}}
\end{center}
\caption{
The Fe\,K$\alpha$ line flux expected from the spherical cloud, 
as a function of viewing angle, $\theta$,
defined to be zero looking through the absorber towards the
ionising source, not including 
the effects of Compton scattering (solid curve).
Also shown is the
mean line flux averaged over viewing angles assuming an isotropic distribution
of clouds
(lower solid horizontal line) and the result from the calculation of R09 (upper
line).  The dashed curve shows the results for the calculation including Compton scattering.
}
\label{fig:lineemission}
\end{figure}

To place limits on the absorber covering factor
in MCG--6-30-15, R09 approximately fit
the {\em Suzaku} data, finding that an absorbing zone of cold gas with
column density $2 \times 10^{24}$\,cm$^{-2}$ and covering fraction
$C_f \simeq 0.35$ can reproduce the observed hard X-ray excess.
We define $C_f$ to be the fraction of the source covered by the absorber
as seen by the observer, and the global covering factor $C_g$ to be the fraction 
of the sky covered by the absorber as seen from the source.
R09 estimated that the zone absorbs
$7.3 \times 10^{-4}$\,photons\,s$^{-1}$\,cm$^{-2}$ from the incident spectrum 
seen by an observer in the rest-frame energy range $7.08-20$\,keV.
They assumed all these photons are absorbed by the Fe\,K
bound-free transition, with Fe\,K 
fluorescent yield 0.347, leading to an
expected Fe\,K (not K$\alpha$) line flux of $2.54 \times 10^{-4}$\,photons\,s$^{-1}$\,cm$^{-2}$.
The calculation is inadequate for two reasons.
\begin{list}{}
  { \setlength{\itemsep}{0pt}
     \setlength{\parsep}{1pt}
     \setlength{\topsep}{1.75pt}
     \setlength{\partopsep}{0pt}
     \setlength{\leftmargin}{0.em}
     \setlength{\labelwidth}{0.em}
     \setlength{\labelsep}{0.2em} }
\item[(i)] Near the Fe\,K absorption edge, that transition only accounts for a fraction 0.51 
of the total photoelectric opacity in low ionisation material \citep{verner96a},
causing R09 to have overestimated the line flux by a factor about 2.
\item[(ii)] Line photons produced within the absorber are 
absorbed as they pass through it.  
The R09 absorber has optical depth at Fe\,K$\alpha$
$\tau \simeq 3.5$, given by the {\sc phabs} function in {\sc xspec}
\citep{arnaud96a}, so reabsorption of line photons cannot be neglected.
\end{list}
In effect, the R09 calculation assumes that the only source
of opacity is the Fe\,K bound-free transition, ignoring the opacity from
other ions and from the outer shells of Fe.
To find the effect of these omissions we calculated the expected line
emission from a spherical cloud, of constant density, absorbing photons from
a distant ionising source.
The gas was assumed to be cold, as in R09, with solar abundances \citep{anders89a} 
and cross-sections of \citet{verner96a}.
For comparison with R09, we first calculate 
the line emission by integrating over the volume of the sphere,
taking account of the inward and outward optical depths as a function of viewing
angle, omitting the effects of Compton scattering.
Fig.\,\ref{fig:lineemission} shows the results 
for a sphere with mean column density $2\times 10^{24}$\,cm$^{-2}$,
illuminated by
a power-law of photon index $\Gamma=2.2$ and normalisation that would be seen by the
observer in the absence of absorption of 
1\,photon\,s$^{-1}$cm$^{-2}$keV$^{-1}$ at 1\,keV.
As the cloud does
not surround the ionising source, the line flux in Fig.\,\ref{fig:lineemission}
should be multiplied by the global covering factor of the cloud.
Also shown is the expected value as calculated by R09: i.e.
measuring the flux removed in the energy range $7.08-20$\,keV and
multiplying by a fluorescent yield of 0.347. 
For a view through the absorber,
the line flux is a factor 11 lower than predicted by R09.
Even viewing the front illuminated face, when line photons
have the highest escape probability, the line flux is a factor 3.7 lower
than predicted by R09. 
The factor by which the R09 calculation overpredicts the line flux depends only weakly
on $\Gamma$.

Suppression of the line flux by photoelectric absorption is inevitable:
if there is sufficient column density to produce detectable line emission there
must also be sufficient opacity to absorb line photons.  However,
the assumed geometry of a system of clouds affects the predictions.
If they isotropically surround the source, 
we can calculate an upper limit to 
the total line flux expected from the
ensemble, averaging over viewing angles, if we neglect ``clouds covering clouds''.
This flux is also shown on Fig.\,\ref{fig:lineemission}
and is a factor 5.7 lower than the R09 prediction. If the
cloud-on-cloud covering factor of the source were 0.35, 
the line emission would be 
reduced by about this factor.  The absorbing system likely has a more
complex geometry, 
but, for any distribution, significantly less line emission is
expected than predicted by R09.  

However, such a high column-density absorber has an optical depth to Compton scattering
$\tau_C \simeq 1.6$, so Fig.\,\ref{fig:lineemission} also shows the line emission predicted
by a Monte Carlo calculation that includes Compton scattering (see also 
\citealt{nandra94b} and \citealt{murphy09a}).  For comparison, the 
front-illuminated equivalent width against the illuminating
continuum is 220\,eV, close to the value, scaled to full covering, of
280\,eV for reflection from an optically-thick disk \citep{george91a}.
For a given illuminating continuum, the line emission is lower than the
case without Compton scattering.  Fitting models to data becomes
more dependent on absorber geometry, however, because the transmitted continuum flux
is attenuated by Compton scattering, but
Compton-scattered light from other clouds enters our line of sight
to compensate: there is scattered continuum whose amplitude is closely
linked to the scattered line flux.  A full calculation of the
expected line flux should take into account the effects of Compton
scattering from the ensemble of clouds, with some assumed geometry.

\section{The covering factor}\label{sec:cov}
Before proceeding, we should consider what is meant
by the global covering factor, $C_g$, in the case 
where the absorber has a 
complex structure. Suppose a wind from the accretion
disk subtends solid angle $\Omega$ at the ionising source, but that within the
wind, absorbing material is fragmented into many clumps that are much smaller
than the projected size of the emitting source: 
the absorbing clumps might themselves cover a 
fraction $f$ of the wind.  In this case the expected line
luminosity $L_{\rm line} \propto \Omega f$. But 
the probability that a randomly-oriented observer would lie on a
sightline passing through the wind is $p \simeq \Omega/4\pi$, without the $f$ factor,
and all such sightlines would see the source covered by the fraction $C_f = f$.
In the R09 model, $C_f=0.35$ and R09 would infer $C_{g} = \Omega f/4\pi \simeq 0.035$,
but in this case the {\em wind} covering factor would be 
$C_{g} = \Omega/4\pi \simeq 0.1$, significantly higher than
the covering factor inferred for the individual clumps within the wind.  Realistic,
complex winds are unlikely to be described by just two numbers, further adding
to the inherent uncertainty in sightline probability.

In the case where the absorbers are part of a disk wind, we could reasonably
expect the accretion disk to extend over at least the same radii as the wind material.  In
this case an observer only sees half of the line-emitting material, 
the other half being obscured 
by the accretion disk \citep{turner09b}.  
If we make the standard assumption that the material is exposed to the
same ionising intensity that the observer infers, 
this occultation introduces up to 
a further factor 2 into the estimate of the wind covering factor
irrespective of whether or not the source itself is occulted.

\section{The MTR model}\label{sec:mtr}
The consequence of such large factors missing from the R09
estimate is that the covering factor deduced could rise from 0.035 to
$\ga 0.5$.
However, R09 did not carry out any calculations
for the actual model fitted to the MCG--6-30-15 data by MTR.  
This has ionised absorbing zones, which might be expected to
increase the line flux for a given high-energy excess, 
as the relative importance of Fe\,K absorption is
expected to increase and the fluorescent yield around Fe\,XIX is higher
than for lower ions \citep[e.g.][]{kallman04a}. On the other
hand, the column densities required were significantly
lower than assumed by R09.  We start by
considering the line emission expected from the MTR model.

There were two key zones, one with $N_H \simeq 4 \times 10^{22}$\,cm$^{-2}$ 
partially-covering the primary X-ray source, designated ``zone 5'',
and a second with 
$N_H \simeq 5 \times 10^{23}$\,cm$^{-2}$ covering only a component of 
reflected emission, designated ``zone 4''.
Note that MTR's first
three absorbing zones, labelled 1--3, are low opacity columns previously discussed by 
\citet{lee01a}, \citet{turner03a, turner04b} and \citet{young05a}.
These produce insignificant line emission.
Although the ``3+2'' zone model requires the introduction of appropriate
free parameters in the fit, zones 1--3 are clearly required
by the presence of narrow absorption lines in the {\em XMM-Newton} and
{\em Chandra} grating data and cannot be ignored, particularly if we wish to model
the entire source spectrum, not a selected sub-region.  The total number of free parameters
required is fewer than in other analyses \citep[e.g.][]{brenneman06a}
and yet the models provide a good description of the entire 0.5-50\,keV spectrum 
across multiple datasets, tested against the full observed range of spectral states.
There is substantial additional information in the spectral variability that
is missed by fitting only to the mean spectrum.

Consider first zone\,5.  This low column is primarily responsible for the
spectral curvature in the ``red wing'' at 2--6\,keV
(see Fig.\,4 of MTR).  Absorption models were generated using
{\sc xstar} \citep{kallman01a, kallman04a} 2.1ln11 and we estimated
the expected line strength from the {\sc xstar} emission-line table.
The {\sc xstar} line
luminosities are those expected from a source with 1--1000\,Ry luminosity
$10^{38}$\,erg\,s$^{-1}$ at a distance of 1\,kpc: these values were scaled to the
amplitude of a power-law continuum integrated over that energy range, yielding
a normalisation factor 0.01695 for a power-law with photon index $\Gamma=2.2$ and 
amplitude 1\,photon\,s$^{-1}$\,cm$^{-2}$\,keV$^{-1}$ at 1\,keV.
As the optical depth at Fe\,K$\alpha$ is only 0.05 in this zone, 
the result is independent of viewing angle.
For $C_g=0.35$,
we find a predicted Fe\,K$\alpha$ line flux of $2.9 \times 10^{-6}$\,photons\,s$^{-1}$\,cm$^{-2}$,
which does not provide any constraint on
the global covering factor of this zone, as the predicted flux is
a factor 8 below the observed line flux of $2.5 \times 10^{-5}$\,photons\,s$^{-1}$\,cm$^{-2}$
(R09).

Zone\,4 of MTR had column density of $5.49 \times 10^{23}$\,cm$^{-2}$
and ionisation parameter $\log\xi = 1.94$, covering reflected emission
modelled by {\sc reflionx} \citep{ross05a}. {\sc reflionx} includes line emission, 
but calculating the line luminosity from the absorbing
zone is more problematic, as the geometry is unknown, 
and the reflector line emission itself is a function of orientation, not modelled in
{\sc reflionx}.  As the optical depth of zone\,4 at Fe\,K$\alpha$ is $\tau \simeq 0.4$, 
its line flux also has some weak orientation dependence.  Thus the following
estimate can only be approximate.
Since the Fe\,K$\alpha$ flux is determined by photons absorbed above the K-edge, 
we use the total high energy continuum flux to estimate the expected
line strength. The total source flux at 20\,keV was 
$f(E)\simeq 0.0015$\,keV\,s$^{-1}$cm$^{-2}$keV$^{-1}$: 
extrapolating over 1--1000\,Ry
assuming $\Gamma=2.2$ and using the zone\,4 {\sc xstar} table
we find an expected Fe\,K$\alpha$ line flux of $5\times 10^{-6}$\,photons\,s$^{-1}$cm$^{-2}$ assuming
full covering.  This is a factor 5 below the observed flux and does not
strongly constrain the MTR model.

While full covering of optically-thick material is not physical, 
a high covering is needed 
to obtain sufficient reflected flux: MTR 
reported that the nominal reflection factor with respect
to reflection from a disk subtending 2$\pi$\,sr is $R\simeq 1.7$, and they
suggested that, instead, 
further highly absorbed zones are likely responsible for a greater proportion
of the hard excess than in the baseline model.  
A further problem for the baseline MTR model is that the ionised absorber zone 4 may also
produce soft-band O\,{\sc vi - viii} line emission, 
which is not observed in the grating data, implying either that the Fe/O abundance ratio
is enhanced or that the highest column density zones
should have lower ionisation than fitted by MTR.  

\section{A pure absorption model}\label{sec:abs}
In the MTR model, absorbed reflection 
dominates the hard excess, but the composition of this component
is not well constrained by the data.  More recently, evidence
has been found for hard excesses caused by high-column partial-covering absorption,
in 1H\,0419-577 \citep{turner09b}, PDS\,456 \citep{reeves09a} and NGC\,1365 
(Risaliti et al., in preparation), 
so we now investigate an extreme model, discussed by MTR, where all of the
hard excess is produced by absorbing, rather than reflecting, zones.
R09 argue that the line emission from such a model should be
representative of the whole class of models explaining the hard excess, for
the purposes of estimating the covering factor.
\begin{table}
\caption{Absorber model parameters, giving for each zone the column density,
ionisation parameter and mean covering fraction, $\overline{C_f}$.  $C_f$ was variable for
zones 4 \& 5. Confidence intervals are 68\%. Quantities in brackets were fixed in
the {\em Suzaku} analysis.}
\label{table1}
\begin{tabular}{cllll}
\hline
zone & N /$10^{22}$cm$^{-2}$ & log($\xi$/erg\,cm\,s$^{-1}$) & $\overline{C_f}$ \\
\hline
1 & 1.18$ \pm .05$     &  $2.39 \pm .01$  & (1.0) \\
2 & 0.027$ \pm .003$    &  $0.88 \pm .16$  & (1.0) \\
3 & (8.0)       &  (3.95)   & (1.0) \\
4 & 191.$ \pm 30$  &  -      & 0.62 \\
5 & 2.9$ \pm .1$   &  $1.38 \pm .03$  & 0.17 \\
\hline
\end{tabular}
\end{table}

\begin{figure*}
\begin{minipage}{0.48\textwidth}{
\resizebox{77mm}{!}{
\rotatebox{270}{
\includegraphics{suzaku_fluxstates.ps}}}
}
\end{minipage}
\hspace*{3mm}
\begin{minipage}{0.48\textwidth}{
\resizebox{77mm}{!}{
\rotatebox{0}{
\includegraphics{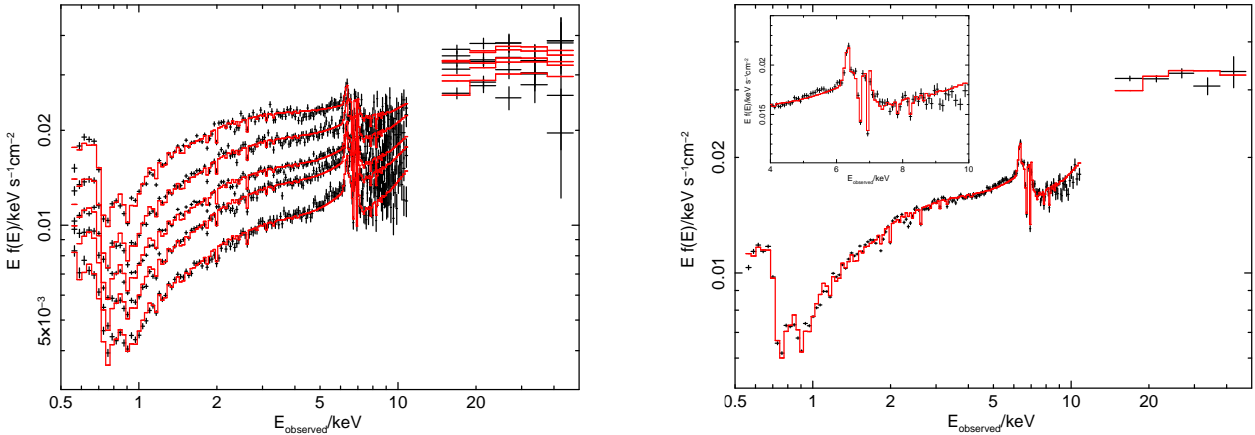}}}
}
\end{minipage}
\caption{
Model fit to the {\em Suzaku} data: (left) divided into five flux states; (right) the mean
spectrum, with insert zoom-in showing the 4--10\,keV region with linear axes.  The model
is shown in units of Ef(E), points with error bars show the ``unfolded'' data.
}
\label{fig:absmodelfit}
\end{figure*}

To test models of MCG--6-30-15, we use the datasets reduced and analysed
by MTR, adopting the same division into flux states.
We emphasise the importance, when $\chi^2$-fitting, of using data binned to match 
the energy-dependent instrument resolution: finer bins
result in significantly less sensitivity to departures from the model 
(section 2.1 of MTR).  Coarser bins are allowed, 
provided that does not erase spectral features.
It is also important to account for systematic errors (section 2.2 of MTR) and
we likewise adopt here a systematic fractional error of 0.03, a likely lower limit
to the calibration uncertainty given cross-instrument comparisons that show 
energy-dependent differences of 5--20\% \citep{stuhlinger07a}. This has a significant 
effect when fitting over the full energy range, as otherwise the goodness-of-fit is
determined almost entirely, but erroneously, 
by the soft band where shot noise uncertainty is very small, 
reducing the accuracy of the model in the crucial Fe\,K region.  Previous
analyses have relied on using data binning that is too fine, and often not including
the soft band in the analysis, to obtain apparently good fits
(e.g. \citealt{miniutti07a}; R09).

To construct the model, we replaced MTR's absorbed reflection (zone 4 absorbing {\sc reflionx}) 
by a single zone of absorption partially-covering the continuum. 
This component has a very hard spectrum and likely comprises a high
column density of low ionisation gas.  We cannot now predict the line luminosity using
{\sc xstar} as the zone is optically-thick at Fe\,K$\alpha$, and we must take 
account of the angular dependence of the line emission.  However, a good
fit to the data may be obtained by assuming this zone to be 
cold gas with transmitted flux modelled by the uniform-density sphere
of section\,\ref{sec:emission},
so we use those results to estimate $C_g$.
In this simple model we ignore Compton scattering, so
the transmitted flux fraction is 
$f(E) = 2 \int_0^1\exp\left[-\frac{3}{2}N\sigma_E\left(1-z^2\right )^{1/2}\right ] z \mathrm{d}z$
where $N$ is the mean column density and $\sigma_E$ is the energy-dependent absorption
cross-section.
Fe\,K$\alpha$, K$\beta$ line emission was added to the model
with line ratio 0.135 \citep{leahy93a}.
We convolved the line emission with a gaussian FWHM 4000\,km\,s$^{-1}$,
consistent with the line width in the {\em Chandra} data.

As we now use {\sc xstar}\,2.1ln11 the parameters of zone\,3 were redetermined
from the {\em Chandra} HEG data, and we found a velocity dispersion of 1000\,km\,s$^{-1}$
best matched the data.  
We include {\sc xstar} line emission from zone\,5 in the fit, although its global covering factor 
is not constrained, so we fix $C_g=0.2$ (including the factors in section\,\ref{sec:cov}).
Fitting simultaneously to the {\em Suzaku} multiple flux states over the
energy range $0.55-50$\,keV, we found a goodness-of-fit of $\chi^2 = 728$ for 
844 degrees of freedom (dof).
Absorber parameters are given in Table\,\ref{table1},
best-fit photon index was $\Gamma=2.20\pm0.003$ and 
Galactic column density $4.2\pm 0.3 \times10^{20}$cm$^{-2}$.
The 0.7\,keV edge depth was degenerate with the zone\,2 parameters
and so was fixed at $\tau = 0.45$.
Fitting to the mean spectrum, with all parameters other than normalisations 
fixed to the multiple flux-state values, resulted in $\chi^2 = 119$ for 170 dof.
There is some indication of a slight model excess in the region
of the Fe\,K edge (Fig.\,\ref{fig:absmodelfit})
but Compton scattering in the high column-density absorber
and some degree of ionisation are both expected to soften the edge.
Strictly, zone\,2 is not required by the {\em Suzaku} data, 
but is required by
the absorption lines in the {\em Chandra} data \citep{lee01a}.
Fitting the same model to the {\em Chandra} data allowing only component 
normalisations to change
resulted in $\chi^2=3538$ for 3346 dof.  Allowing the warm absorber
parameters to vary resulted in $\chi^2=3427$ for 3341 dof, showing good agreement
with the high resolution mean spectrum.
This model is not unique, and does not include any scattered continuum (see below),
but the goodness-of-fit is already
sufficiently good that, statistically, 
no additional components are justified.

The Fe\,K$\alpha$ flux (assumed constant) was 
$2.27 \pm 0.2 \times 10^{-5}$ photon\,s$^{-1}$\,cm$^{-2}$,
the mean continuum flux incident on the absorber was 
$0.0383 \pm 0.0012$\,photons\,s$^{-1}$cm$^{-2}$keV$^{-1}$ at 1\,keV. 
Using the section\,2 line flux
for $N_H = 1.9\times 10^{24}$\,cm$^{-2}$,
we infer a mean global covering factor $C_g \simeq 0.45 \pm .04$
(statistical errors only, 68\% confidence intervals), including the factors in section\,\ref{sec:cov}.

However, as noted in section\,\ref{sec:emission},
the optical depth to Compton scattering in this absorber would be
$\tau_C \simeq 1.5$, so the true incident flux onto this component could
be a factor 5 higher than inferred and 
the deduced mean covering fraction, $\overline{C_f}$,
could be
substantially closer to unity after allowance for scattering.
The hard-band flux would also have a significant contribution
from light scattered into our line of sight from the other absorbers 
around the source (the ``absorbed reflected'' component in the original MTR
model may be a representation of such a Compton-scattered component).
To achieve a better measure of $C_g$
we need to model the spectrum that is scattered into
our line of sight (Miller et al., in preparation).

\section{Discussion}\label{sec:discussion}
Much of the motivation for the work of MTR was the observation that the red wing
and hard excess appear fairly constant, despite large variations
in source brightness at $E<10$\,keV.  If these components are reflected emission then 
their brightness should vary in phase with the illuminating continuum, as the light
travel time between source and inner accretion disk should be negligible.  
A popular explanation for the lack of such variation is the 
``light-bending'' model of \citet{fabian03a}, \citet{miniutti03a} and \citet{miniutti04a}.
\citet{miniutti03a} show for MCG--6-30-15 that if the illuminating source is within about $3-4$\,GM/c$^2$ of the
black hole, and if it varies in height over about $3-8$\,GM/c$^2$, then the effects of light following
geodesics in the spacetime near the black hole mean that the flux 
reaching the distant observer from the source can be arranged to decrease as it approaches the accretion disk,
while the observed flux reflected from the disk may remain approximately constant.
In this picture the source variability is caused by motion of the central
source:
obtaining a relatively
constant reflected intensity requires both the supposition of a compact vertically-moving source and its
careful placement.

The alternative explanation presented here implies that a substantial fraction of the observed
variability is caused by variations in covering fraction of clumpy absorbers.  We suggest that
the absorbers may be a complex multi-component structure, with a wide range of scale-sizes,
possibly part of an accretion disk wind \citep[e.g.][]{proga04a}.  In this picture we
expect the source variability to decrease to higher energy, as found for MCG--6-30-15, 
and that more absorbed AGN should show greater hard X-ray variability than less absorbed 
AGN, as found by \citet{beckmann07a}.
There is expected to be reflection from the most optically-thick absorbers, but if the
intrinsic source variations are low, both the reflection and any line emission 
would likewise be of relatively constant amplitude.  The simple parameterisations 
presented here are likely not sufficient for detailed investigation of the
X-ray spectra, and models of radiative transfer that treat scattering and absorption
in the velocity field of the wind are needed \citep{sim08a}.

\section{Conclusions}
The variability, ``red wing'' and hard excess in the X-ray spectrum of MCG--6-30-15
may all be explained by a model of clumpy absorption.  The observed Fe\,K$\alpha$
line emission may be fluorescent emission from the absorbing material, in which case the
global covering factor of the material is $C_g \simeq 0.45$.  The largest uncertainty in this
estimate is caused by lack of knowledge of the absorber geometry.  In calculating the expected
line emission it is essential to take account of photoelectric absorption of line photons,
neglect of which could result in predictions that are incorrect by factors of order 10.  
Radiative transfer models that include the effect of Compton scattering are required
to reach a full understanding of such systems.  If correct, this model implies that
a large fraction of the observed X-ray variability may be caused by variation in absorber covering
fraction, and that a large fraction of the source $2-10$\,keV
luminosity may be partially obscured even in type\,I AGN.

\vspace*{1ex}
\noindent
{\bf Acknowledgments}.
We are grateful to Tim Kallman for providing and updating {\sc xstar}. 
Rapid production of 
{\sc xstar} tables was possible using {\sc pvm\_xstar} \citep{noble09a}.
TJT acknowledges NASA grant NNX08AJ41G.

\bibliographystyle{mn2e}
\bibliography{xray}

\label{lastpage}

\end{document}